\documentclass[preprint,12pt]{elsarticle}

\usepackage{amssymb}
\usepackage{nicefrac}

\usepackage{lineno}
\usepackage[hidelinks]{hyperref}
\hypersetup{colorlinks,bookmarksopen,linkcolor=blue,citecolor=blue,urlcolor=blue}
\usepackage{xcolor}

\journal{Elsevier}

\usepackage[english]{babel}
\usepackage[utf8x]{inputenc}
\usepackage[T1]{fontenc}
\usepackage{bm}

\usepackage{pgfplotstable}
\usepackage{pgfplots}
\usepackage{tikz}
\usepackage{stanli}
\usepackage{amsmath}
\usepackage{amsfonts}
\usepackage{url}
\pgfplotsset{compat = 1.15}
\usepackage{graphicx}
\usepackage{color}
\usepackage{geometry}
\usepackage{setspace}
\usepackage{siunitx}
\usepackage{multirow}
\usepackage{graphicx}
\usepackage{changepage}
\usepackage{chngcntr}

\usetikzlibrary{external}
\usetikzlibrary{angles,quotes}

\usepackage{rotating}

\usepackage{tabularray}
\usepackage{placeins}

\usepackage{tabularray} 
\UseTblrLibrary{booktabs} 

\usepackage{pgfplots}
\usepackage{pgfplotstable}
\pgfplotsset{compat=newest}

\usepgfplotslibrary{external}
\begin{document}
\emergencystretch 3em

\begin{frontmatter}



\title{Deformation gradient averaging regularization for third medium contact}


\author[ctumech]{Ondřej Faltus\corref{cor1}}
\ead{ondrej.faltus@cvut.cz}
\cortext[cor1]{Corresponding author}
\author[ctumech]{Marco Amato}
\author[ctumech,casita]{Martin Horák}

\affiliation[ctumech]{
            organization={Department of Mechanics, Faculty of Civil Engineering, Czech Technical University in Prague},
            addressline={Thákurova 7}, 
            postcode={166 29},
            city={Prague 6},
            country={Czech Republic}}

\affiliation[casita]{
            organization={Institute of Information Theory and Automation, Czech Academy of Sciences},
            addressline={Pod Vodárenskou věží 4}, 
            postcode={182 00},
            city={Prague 8},
            country={Czech Republic}}


\begin{abstract}
The third medium contact method has recently come into popularity as an alternative to traditional contact methods in contexts where search for contact boundaries is problematic, i.e. topology optimization. To enforce the contact constraints, it relies on a fictitious compliant material occupying the void space. In finite strain setting, this necessitates regularization techniques to improve the behavior of the third medium material. A number of existing models rely on penalization of locally computed second gradients of displacements, either through direct calculation on second-order elements or through additional degrees of freedom. Here we propose an alternative approach using element-wise deformation gradient averaging to effectively penalize spatial variations of the deformation gradient, together with linear elastic term enforcing constant third medium stiffness. Our approach enables the use of first-order finite element formulations without any additional degrees of freedom and is therefore easy to implement. We demonstrate the robustness of the proposed method on several well-established benchmarks.
\end{abstract}

\begin{keyword}
Contact mechanics \sep Third medium method \sep Deformation-gradient averaging \sep Regularization
\end{keyword}

\end{frontmatter}


\section{Introduction}

Advanced material and structure designs, as well as improved computational methods necessary to reliably simulate their behavior, are increasingly prevalent in modern materials science~\cite{Ren2018,yu2018metamatreview,liu2026softBiomimeticUnderwaterReview,elishakoff2021galerkin,elishakoff2021flutter}. In this context, computational contact mechanics approaches are becoming increasingly important to capture the highly nonlinear behavior that arises in such problems.

The enforcement of contact constraints in finite element simulations can be achieved using a variety of formulations. Classical approaches include the penalty, Lagrange multiplier, and augmented Lagrange methods~\cite{laursen2003contactbook, Wriggers2006contactbook, Konyukhov2015, simo1985lmcontact}. These formulations are typically combined with different contact discretization strategies, where contact search algorithms detect potential interacting surfaces and evaluate the contact gap. Common discretizations include node-to-node, node-to-segment, segment-to-segment, and mortar methods~\cite{Wriggers2006contactbook,Konyukhov2015,Zavarise2009improvedNTS, Zavarise1998s2s,fischer2005frictionlessmortar}.

The third medium contact method is different from these traditional approaches. It enforces contact constraints through the introduction of a fictitious material occupying the void space between the contacting bodies. The parameters of this material are chosen such that it is extremely compliant in the pre-contact phase in order not to significantly influence the solution, yet it stiffens when compressed to near-zero volume, enforcing contact while allowing a small positive gap to remain. This positive gap is analogous to the negative gap present in the penalty-based contact formulations. The main advantage of the third medium contact method is that it avoids the need for a contact search algorithm. This is especially convenient in contexts where contact search cannot be readily performed, e.g., in topology optimization~\cite{Bluhm2021contact}.

The third medium method was first proposed by Wriggers~et~al.~\cite{Wriggers2013thirdmedium}. In this early iteration, the third medium was modeled using  an anisotropic material model whose anisotropy direction is determined by the contact normal, identified from the direction of principal stretches. Bog~et~al.~\cite{bog2015tmcontactwithbarrier} later introduced a barrier method to control the third medium simulation and Kruse~et~al.~\cite{Kruse2018isogeomthirdmedium} extended the approach to isogeometric analysis.

A surge of third medium popularity has come with the work of Bluhm~et~al.~\cite{Bluhm2021contact}, who applied the third medium method to topology optimization simulations. In this context they effectively utilized the already discretized void space to enforce contact constraints. However, due to the high compliance of the third medium, the medium tended to deform excessively in the pre-contact phase. To address this problem, the strain energy of the third medium was enhanced by introducing an additional regularization term involving locally computed second gradients of the displacement field. This modification stabilized the behavior of the third medium and enabled robust finite-strain simulations. The approach was further improved in the follow-up work of Frederiksen~et~al.~\cite{Frederiksen2023topologyopt}, and later extended to thermal contact~\cite{dalklint2025thermal3M} and Coulomb friction~\cite{frederiksen2024friction}.

One of the disadvantages of this otherwise elegant regularization approach is that it penalizes more deformation modes of the third medium elements than strictly necessary. In a subsequent study, a reduced regularization form was proposed that avoids penalizing of bending and quadratic compression modes~\cite{frederiksen2025huhululu}.

The approach was later adapted and extended for pneumatically actuated structures~\cite{Faltus20243M}. In that work, the regularization term was replaced by a more permissive formulation that penalizes gradients of the logarithm of rotations, approximated through incremental accumulation of the spin tensor. This regularization was later improved by Wriggers~et~al.~\cite{wriggers20253Mimprovedrotations}, who introduced a direct computation of the rotation tensor from the rotation angle, which is valid in two-dimensional formulations. In this contribution, an approach based on first-order finite elements with additional degrees of freedom was also proposed. Previously, the regularization terms required second-order elements for the third medium. Later, Wriggers~et~al.~\cite{wriggers20253Mlinear} extended this work by introducing a universal regularization term valid also in three dimensions, which penalizes the skew-symmetric part of the deformation gradient. Once again, first-order element formulation becomes possible through the introduction of additional degrees of freedom. The discretization of these degrees of freedom has recently been improved by von~Zabiensky~et~al.~\cite{vonzabiensky2026neighboredelement3M}. For two-dimensional problems, the third medium contact approach has also recently been extended to the virtual element method (VEM) by Xu and Wriggers~\cite{xu2026Wriggers3MVEM}.

In this article, we present a new regularization method which completely avoids the use of second gradients of displacement. Instead, it relies on a regularization term penalizing deviations of the deformation gradient from its element-centroid value. This results in a very simple yet robust formulation and enables the use of first-order elements without introducing any additional degrees of freedom. In Section~\ref{sec:model}, we present the formulation of the proposed regularization term, and in Section~\ref{sec:examples} we demonstrate its applicability through several numerical examples. Finally, Section~\ref{sec:conclusions} summarizes the main findings and provides concluding remarks.

\section{Third medium model}
\label{sec:model}

\subsection{Combined hyperelastic and linear elastic terms}
\label{sec:constantstiffnessterm}

Traditionally, the neo-Hookean material law~\cite{rivlin1948neohookean} adapted for compressible materials is used as the basis for the third medium models~\cite{Bluhm2021contact,Faltus20243M,wriggers20253Mimprovedrotations,wriggers20253Mlinear}. It is defined by the strain energy density in the form

\begin{equation}
    W_\mathrm{NH} = \frac{1}{2}\kappa_\mathrm{vol} \ln^2{J} + \frac{1}{2}\kappa_\mathrm{iso} (J^{-2/3}I_1 - 3)^2 
\label{eq:neohookenergy}
\end{equation}

\noindent where $J=\det{\bm{F}}$ is the Jacobian of the deformation gradient~$\bm{F}$, and $I_1 = \mathrm{tr}(\bm{F}^T\bm{F})$ is the first invariant of the right Cauchy-Green deformation tensor. The material parameters of the volumetric part $\kappa_\mathrm{vol}$ and of the isochoric part $\kappa_\mathrm{iso}$ are typically chosen so that the third medium has negligible stiffness and thus only a negligible influence on the solution unless it is extremely compressed ($J \to 0$). A typical choice is to relate them to the parameters of the bulk material, e.g., by setting

\begin{equation}
\label{eq:neohookcontactstiffness}
    \kappa_\mathrm{vol} = \kappa_\mathrm{iso} = \gamma_\mathrm{c} \kappa_\mathrm{char}
\end{equation}

\noindent where $\gamma_\mathrm{c}$ is the relative third medium stiffness parameter, typically in the range of $\langle 10^{-10}, 10^{-6}\rangle$, and $\kappa_\mathrm{char}$ is the characteristic bulk modulus of the bulk material.

This material model allows for a smooth transition between the hyperelastic bulk material and third medium in topology optimization applications~\cite{Bluhm2021contact,Frederiksen2023topologyopt,dalklint2025thermal3M,frederiksen2025huhululu}, while fulfilling the requirements for the desired behavior of the third medium. Under extreme compression ($J \to 0$), the logarithmic volumetric term in the strain energy density~(\ref{eq:neohookenergy}) approaches infinity and prevents penetration. It has been shown in Faltus~et~al.~\cite{Faltus20243M} that, under plane strain assumption, this property is satisfied by the isochoric term as well, and thus the volumetric term can even be omitted.

Here we adopt a linear elastic behavior for the third medium in the pre-contact phase in order to improve its response prior to contact, following the idea of Wang et al.~\cite{Wang2014tointerpolationreg}, who modeled void regions using a linear elastic formulation to prevent excessive distortion of low-stiffness elements and thereby improve the robustness of geometrically nonlinear simulations in the topology optimization context. In contrast to our current formulation, no explicit contact term was introduced in that work. Moreover, the present approach can be interpreted as introducing a constant third medium stiffness (in the $\bm{P}$–$\bm{F}$ formulation), i.e., the stiffness does not increase with deformation except for the asymptotically increasing response under extreme compression, which is captured by the contact term.

The small-strain linear elastic material law is defined by the strain energy density function

\begin{equation}
\label{eq:smallstrainenergydensity}
    W_\mathrm{ss} = \frac{1}{2}\bm{\varepsilon} : \mathbf{D}^\mathrm{e} : \bm{\varepsilon}
\end{equation}

\noindent where $\bm{\varepsilon}$ is the engineering strain defined as the symmetric gradient of the displacement field, $\bm{\varepsilon} = \nabla_\mathrm{s}\bm{u}$, and $\mathbf{D}^\mathrm{e}$ is the small-strain elasticity tensor.

The small-strain stress tensor $\bm{\sigma}$ is then obtained as

\begin{equation}
    \bm{\sigma} = \frac{\partial W_\mathrm{ss}}{\partial \bm{\varepsilon}} = \mathbf{D}^\mathrm{e} : \bm{\varepsilon}
\label{eq:smallstrainlaw}
\end{equation}

In a large-strain context, the strain energy density~(\ref{eq:smallstrainenergydensity}) can be expressed in terms of the deformation gradient $\bm{F}$:

\begin{equation}
    W_\mathrm{ss} = \frac{1}{2}\left(\bm{F}-\bm{I}\right) : \mathbf{D}^\mathrm{e} : \left(\bm{F}-\bm{I}\right)
\end{equation}

\noindent where $\bm{I}$ is the second-order identity tensor. This formulation exploits the minor symmetries of $\mathbf{D}^\mathrm{e}$ to replace the symmetric displacement field gradient $\bm{\varepsilon} = \nabla_\mathrm{s} \bm{u}$ with the general gradient $\left(\bm{F}-\bm{I}\right) = \nabla \bm{u}$ without affecting the resulting energy density. A detailed discussion together with implementation details is provided in \ref{app:isole_details}.

The validity of this material law for physical materials relies on the standard small-strain assumptions; in a large-strain context it results in a non-objective formulation that produces non-zero energy under pure rotations. Nevertheless, the third medium is fictitious and we therefore adopt this formulation despite this limitation. Given that the stiffness of the third medium is typically at least six orders of magnitude smaller than the stiffness of the bulk material, any spurious energy generated remains negligible unless the third medium undergoes extremely large rotations. We remark that due to the existence of an energy potential, this approach preserves the symmetry of the global stiffness tangent.

The isotropic small-strain stiffness tensor is fully defined by the two parameters $E$, the Young's modulus, and $\nu$, the Poisson's ratio. For the third medium material stiffness tensor $\mathbf{D}^{\mathrm{3M}}$, we choose Poisson's ratio of $\nu = 0$ and a Young's modulus scaled with respect to the bulk material:

\begin{equation}
    E_\mathrm{3M} = \gamma_\mathrm{e} E_\mathrm{char}
\end{equation}

\noindent where $\gamma_\mathrm{e}$ is the relative stiffness parameter of the third medium, which is not necessarily identical to its counterpart $\gamma_\mathrm{c}$ from Equation~(\ref{eq:neohookcontactstiffness}), and $E_\mathrm{char}$ is the characteristic Young's modulus of the bulk material. For hyperelastic bulk materials, this quantity can be taken as the Young's modulus obtained by linearization around the reference configuration.

\subsection{Strain averaging regularization term}
\label{sec:regularization}

Since the work of Bluhm~et~al.~\cite{Bluhm2021contact} most third medium formulations employ some form of second-gradient regularization to improve the pre-contact behavior of the third medium. This is most easily achieved by augmenting the strain energy density of the third medium with a regularization term involving second displacement gradients, typically written as

\begin{equation}
    W_{\nabla F} = \frac{1}{2} \gamma \kappa_{\nabla\bm{F}} \left( \nabla \bm{F} \;\vdots\; \nabla \bm{F}\right)
\end{equation}

\noindent where $\kappa_{\nabla\bm{F}}$ is a material parameter and $\nabla\bm{A}$ denotes the material gradient operator applied to the second-order tensor $\bm{A}$, i.e., $(\nabla\bm{A})_{ijm} = \partial A_{ij}/\partial X_m$, where $\bm{X}$ is the material coordinate vector.

Other works extend this idea by penalizing alternative, less restrictive measures or combinations thereof. Among the regularization terms reported in literature, we find

\begin{itemize}
    \item $\nabla\bm{Q}$, the gradient of a rotation tensor incrementally calculated from material spin~\cite{Faltus20243M},
    \item $\nabla\bm{R}$, the gradient of the true rotation tensor in 2D~\cite{wriggers20253Mimprovedrotations},
    \item $\nabla\varphi$, the gradient of the 2D rotation angle~\cite{wriggers20253Mimprovedrotations},
    \item $\mathrm{Div}\bm{F}$, the divergence of the deformation gradient tensor~\cite{frederiksen2025huhululu},
    \item $\nabla\bm{F}_\mathrm{skw}$, the gradient of the skew-symmetric part of the deformation gradient tensor~\cite{wriggers20253Mlinear}, and
    \item $\nabla J$, the gradient of the deformation gradient Jacobian~\cite{Faltus20243M,wriggers20253Mimprovedrotations}.
\end{itemize}

All of these terms contain the second gradients of displacement and in a physical material would thus require a second-gradient continuum formulation. The fictitious nature of the third medium, however, allows this requirement to be relaxed and the penalization can instead be applied only at the element level. Either finite elements with a quadratic formulation must then be used~\cite{Bluhm2021contact,Faltus20243M,wriggers20253Mimprovedrotations,frederiksen2025huhululu}, or first-order elements must be enhanced with additional degrees of freedom~\cite{wriggers20253Mimprovedrotations, wriggers20253Mlinear, vonzabiensky2026neighboredelement3M}. Here we aim to overcome the limitations of both approaches by introducing a simpler formulation that allows the direct use of first-order finite elements without additional degrees of freedom, thereby eliminating the need to compute second gradients.

The penalization of the gradient of the deformation gradient tensor (or another deformation measure), when computed only element-wise, effectively allows one to control the level of uniformity of this tensor across the element, or, more precisely in the finite element setting, across its Gauss integration points. However, achieving this effect does not require the explicit computation of gradients. A similar result can be obtained by penalizing deviations of the deformation gradient evaluated at the Gauss integration points from its element-wise average value, taken here as the deformation gradient at the element centroid. In other words, we penalize differences between the deformation gradient at the integration points and the representative deformation gradient of the element. To this end, we introduce an additional strain energy density of the form
 
\begin{equation}
    W_{\bar{F}} = \dfrac{1}{2}\gamma\kappa_{\bar{F}} \|\bm{F}-\bar{\bm{F}}\|^2
\label{eq:barFenergy}
\end{equation}

\noindent where $\kappa_{\bar{F}}$ is a penalty parameter, $\bar{\bm{F}}$ denotes the deformation gradient evaluated at the element centroid, and $|\bm{a}|$ represents the Frobenius norm of a tensor $\bm{a}$. With this formulation, the level of uniformity of the deformation gradient within the finite element can be controlled through the choice of the parameter $\kappa_{\bar{F}}$. Larger values of $\kappa_{\bar{F}}$ enforce stronger uniformity of the deformation gradient across the element, while smaller values allow greater variation. The simple modification of the finite element procedure required to evaluate $\bar{\bm{F}}$ is described in \ref{app:fbar_details}.

Within this contribution we limit ourselves, for simplicity, to the averaging of $\bm{F}$ according to Equation~(\ref{eq:barFenergy}). The same approach, nevertheless, would be possible for the penalization of the other measures quoted above.

Two points should be noted here. Firstly, due to the element-wise averaging nature of this penalization, it requires elements with multiple integration points with differing values of deformation gradient. First-order quadrilateral elements with four integration points and bilinear shape functions are therefore particularly suitable. Secondly, the strength of this regularization depends on the spacing of the integration points, and thus on the mesh size. We investigate this relationship in the numerical example in Section~\ref{sec:cshape}.

\subsection{Model summary}

The full proposed material model for the third medium consists of three parts, namely

\begin{enumerate}
    \item the contact term based on the volumetric part of the neo-Hookean material law~(\ref{eq:neohookenergy}). The isochoric part is dropped to ensure that this part only reacts to the volumetric compression of the third medium and does not interfere with the behavior of
    \item the constant stiffness term based on the linear elastic law~(\ref{eq:smallstrainlaw}) to ensure constant behavior especially in the pre-contact phase, and
    \item an additional regularization term based on deformation averaging.
\end{enumerate}

Those terms can be collected in the full strain energy density functional

\begin{equation}
    W_\mathrm{3M} = \frac{1}{2}\kappa_\mathrm{vol} \ln^2{J} + \frac{1}{2}\left(\bm{F}-\bm{I}\right) : \mathbf{D}^\mathrm{3M} : \left(\bm{F}-\bm{I}\right) + \dfrac{1}{2}\kappa_{\bar{F}} \|\bm{F}-\bar{\bm{F}}\|^2
\end{equation}

\section{Numerical examples}
\label{sec:examples}

To calculate the following numerical examples, the proposed third medium model has been implemented into the open-source finite element software OOFEM~\cite{patzak2001oofem,patzak2012oofem1,patzak2012oofem2}. Unless stated otherwise, the simulations are performed under the plane strain assumption using bilinear four-node quadrilateral finite elements with the Lobatto integration rule. To improve integration accuracy, 9 instead of 4 integration points are used for the elements representing the bulk material. The Newton-Raphson solver has been augmented with line search~\cite{Bonet2016nonlinearcontinuumfem} and adaptive time stepping algorithms. In simulations involving stability (Section~\ref{sec:beaminbox}) the modified Cholesky decomposition algorithm has been introduced to steer the solver towards stable solution branches~\cite{Cheng1998modifiedcholesky, Nocedal2006numericaloptimization}. Unless stated otherwise, the bulk material is described by the compressible neo-Hookean model, see Equation~(\ref{eq:neohookenergy}), with the parameters $\kappa_\mathrm{vol} = 10^6\;\si{\pascal}$ and $\kappa_\mathrm{iso} = 0.214\cdot10^6\;\si{\pascal}$, which correspond in the small-strain limit to $E = 3\cdot10^{5}\;\si{\pascal}$ and $\nu = 0.4$.

The data for the numerical simulations presented here are available in an associated data repository~\cite{Faltus2026improved3MZenodo}.

\subsection{C-shape benchmark}
\label{sec:cshape}

\begin{figure}[p]
    \centering
    \includegraphics[width=\textwidth]{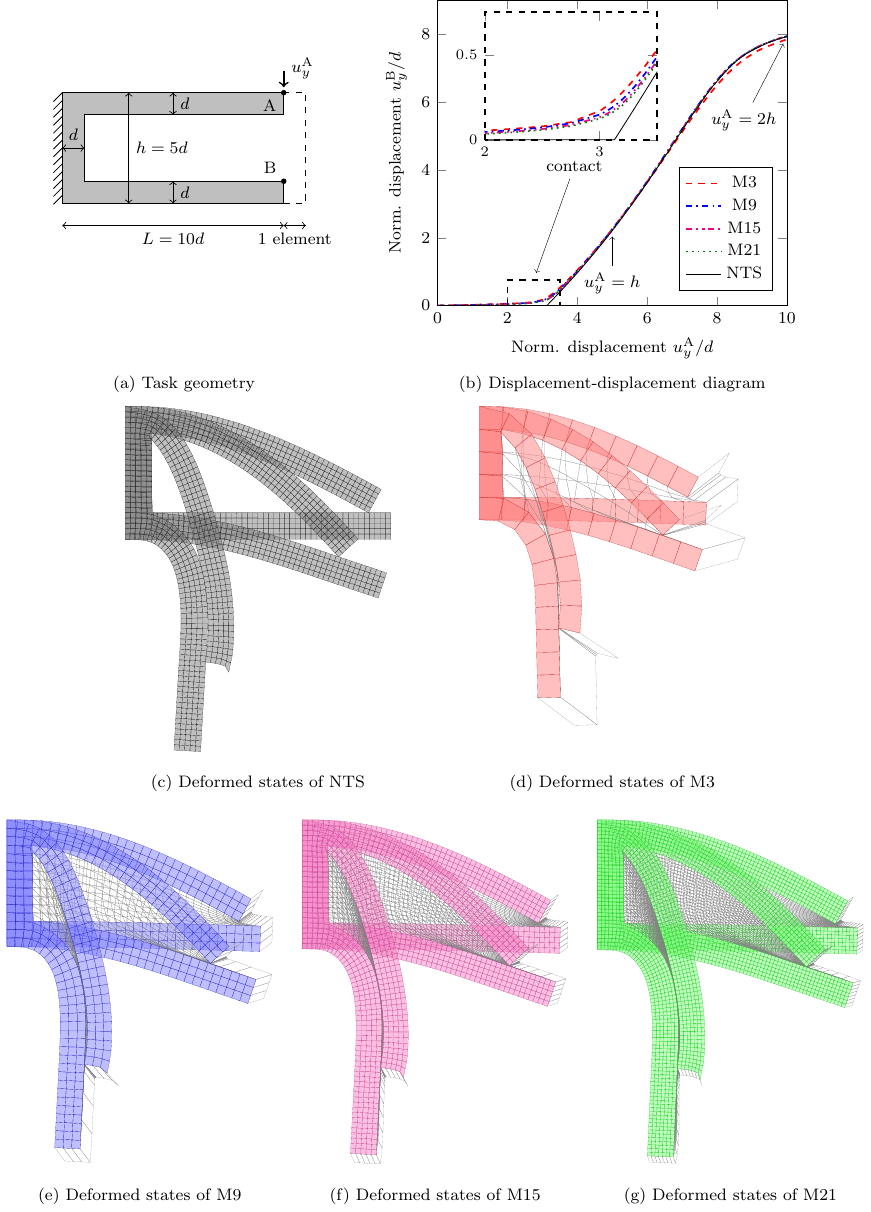}
    \caption{The C-shape benchmark with different mesh refinement. (a) Task geometry. (b) The dependence of the vertical nodal displacement of node $B$ of the bottom cantilever on the prescribed vertical displacement of node $A$ of the top cantilever with a detail of the contact point shown in the inset. Results are shown for the four different meshes from Table~\ref{tab:cshapeparams}, as well as for a traditional node-to-segment double-pass contact solution. (c-g)  Deformed states of the individual meshes at points annotated in the plots, bulk elements are shown in color, whereas third medium elements are shown in gray outline.}
    \label{fig:cshape}
\end{figure}

The classical C-shape benchmark for third medium contact, first introduced by Bluhm~et~al.~\cite{Bluhm2021contact}, is presented here together with a mesh-size study. The task geometry consists of two cantilevers placed one above the other with a third medium mesh between them, see Figure~\ref{fig:cshape}a. The third medium has a free surface, with an extra column of elements added along it. See~\cite{Faltus20243M} for an explanation of why this is necessary. Loading is realized in the form of an applied vertical displacement on the rightmost node of the top cantilever's top surface (node $A$). Figure~\ref{fig:cshape}b then shows the dependence of the displacement of node $B$ on the bottom cantilever on the prescribed displacement of node $A$, demonstrating the onset of contact when the top cantilever closes the gap.

The simulation is conducted on four different third medium meshes with different numbers of elements. Their details are summarized in Table~\ref{tab:cshapeparams}. They are named according to the number of elements arranged vertically in the space between the cantilevers, i.e., in the contact gap. Due to the dependency of the deformation averaging's effectiveness on the spacing between element integration points, the value of the penalty parameter $\kappa_{\bar{F}}$ had to be adjusted with the mesh size (increasing as the element size decreases). The other parameters of the third medium are the same for all meshes, namely $\gamma_\mathrm{c} = \gamma_\mathrm{e} = 10^{-6}$.

Additionally, a standard contact simulation has also been performed for comparison purposes. A double-pass node-to-segment contact strategy was employed, with the contact constraint enforced by the penalty method. The mesh of the bulk material corresponds to the M15 third medium mesh, with 5 elements per beam cross-section height.

\begin{table}[ht]
  \centering
  \begin{tblr}{
    colspec = {>{\bfseries}l|*{4}{X[r]}},
    row{1} = {font=\bfseries, halign=left},
    row{6} = {abovesep=10pt},
    vlines,                        
    hline{1,Z},                    
    hline{2} = {1pt},              
  }
    Mesh & Beam  & Contact gap & $\kappa_{\bar{F}}$ parameter & Gap error \\
    M3     & 1 elem./height  & 3M, 3 elements  & $\SI{200}{\pascal}$  & $6.8 \%$ \\
    M9     & 3 elems./height & 3M, 9 elements  & $\SI{600}{\pascal}$  & $5.9 \%$ \\
    M15    & 5 elems./height & 3M, 15 elements & $\SI{1000}{\pascal}$ & $5.3 \%$ \\
    M21    & 7 elems./height & 3M, 21 elements & $\SI{1400}{\pascal}$ & $5.0 \%$ \\
    NTS    & 5 elems./height & penalty, $10^{6}\;\si{\pascal}$ & -                  &  -       \\
  \end{tblr}
  \caption{Parameters of the different meshes for the C-shape benchmark. Note the $\kappa_{\bar{F}}$ penalty parameter increasing with mesh refinement in the contact gap. The last mesh is a traditional node-to-segment contact mesh with a penalty parameter.  The last column shows the error of contact enforcement measured as the size of the residual gap left at the contact point relative to the initial gap.}
  \label{tab:cshapeparams}
\end{table}

With a linear scaling of the $\kappa_{\bar{F}}$ parameter, the results are comparable for the different mesh sizes. The error in the sharpness of contact enforcement is reported in Table~\ref{tab:cshapeparams}, while the evolution of the nodal displacements of both cantilevers is shown in Figure~\ref{fig:cshape}b. In the zoomed-in inset of Figure~\ref{fig:cshape}b we can see the rounded profile of the curves at the contact point characteristic of the third medium method. The traditional node-to-segment method captures the contact point more clearly, see also the discussion on this topic by Wriggers~et~al.~\cite{wriggers20253Mimprovedrotations}. Selected deformed states of the different meshes at loading levels up to $u^A_{y} = 2h$ are pictured in Figures~\ref{fig:cshape}c\text{-}g. The M3 mesh is the only one for which the results significantly differ from the others. This is, however, a consequence of the coarseness of the mesh with only one linear element across the cross-section height of the bending cantilevers, which leads to an unrealistically stiff response. 

\subsection{Closed box benchmark}

\begin{figure}[t]
    \centering
    \includegraphics[width=\textwidth]{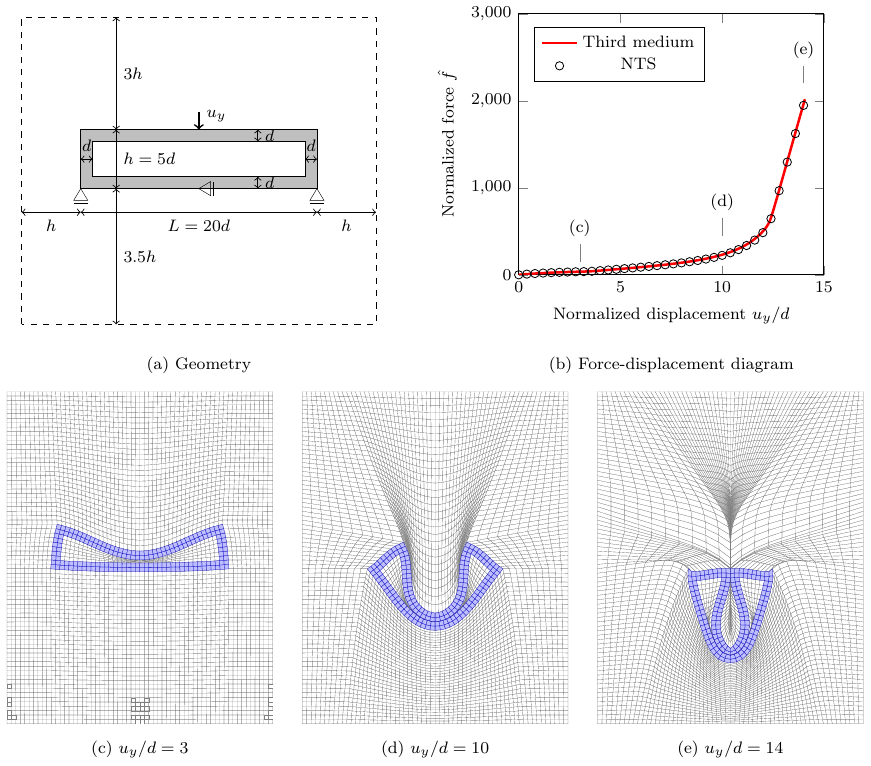}
    \caption{The closed box benchmark with double self-contact. (a) Task geometry (the third medium envelope is not to scale in the vertical direction). (b) Force-displacement diagram. The annotations show the positions corresponding to the different load levels of the displacement $u_y$, for which the deformed mesh configurations are shown in (c-e). Bulk elements are shown in blue, third medium in gray.}
    \label{fig:closedbox}
\end{figure}
Another established benchmark for third medium contact involves self-contact of a closed box under either pneumatic or mechanical loading. Here we extend this benchmark further by simulating displacement loading up to the point where the box bends and self contact is established also between its outer sides. The geometry of the task is presented in Figure~\ref{fig:closedbox}a. The box is supported symmetrically against rigid-body motion and the loading is applied as a prescribed displacement $u_y$ at the center of the top surface. To simulate the outer contact, the entire box has to be enclosed in a third medium mesh of sufficient size. The displacements of its sides are fixed. The parameters of the third medium used in this simulation are $\gamma_\mathrm{c} = 10^{-7}$, $\gamma_\mathrm{e} = 10^{-6}$, and $\kappa_{\bar{F}} = \SI{500}{\pascal}$.

The force-displacement diagram of the task is shown in Figure~\ref{fig:closedbox}b. The displacement is normalized in increments of the beam cross-section height, while the force is shown in a dimensionless representation $\hat{f}$ calculated as

\begin{equation}
    \hat{f} = \frac{fL^3}{Etd^4}
\end{equation}

\noindent where $f$ is the raw reaction force, $E$ is the Young's modulus of the bulk material, $t$ is the out-of-plane thickness, and $L$ and $d$ are parameters of the geometry, see Figure~\ref{fig:closedbox}a. This normalization makes $\hat{f}$ independent of both the task geometry and the bulk material parameters.

Again, we compare with a traditional node-to-segment method here. The settings of the node-to-segment simulation are the same as in the case of the C-shape benchmark in Section~\ref{sec:cshape}. A remarkably good agreement with the third medium approach is reached in the closed box example, as illustrated by the overlaid curve and markers in Figure~\ref{fig:closedbox}b. The deformed states at load levels expressed in multiples of the box thickness $d$ are pictured in Figure~\ref{fig:closedbox}c-e. The third medium approach results are shown, but the deformed mesh of the traditional contact coincides almost perfectly with the bulk elements of the third medium mesh anyway.

\subsection{Constrained buckling and self-contact}
\label{sec:beaminbox}

\begin{figure}[t]
    \centering
    \includegraphics[width=\textwidth]{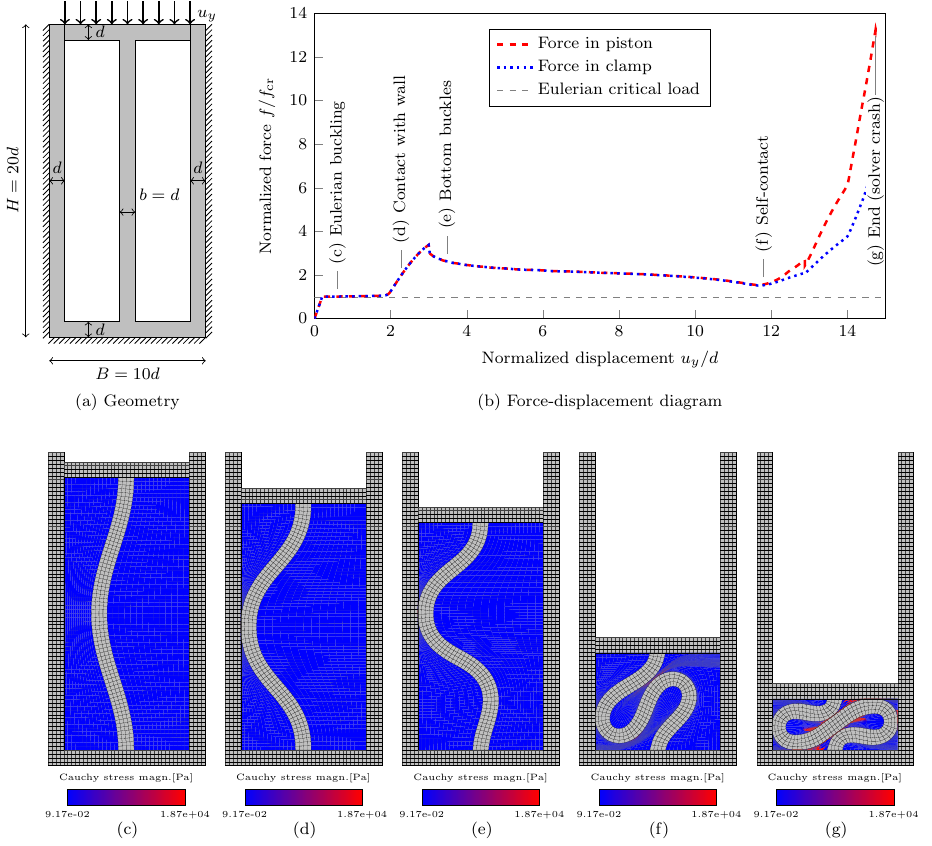}
    \caption{Constrained compression of a double-clamped beam with instability. (a) Task geometry. (b) Force-displacement diagram. The annotations show the positions corresponding to the different load levels of the displacement $u_y$, for which the deformed mesh configurations are shown in (c-g). The colors in the deformed state figures correspond to the magnitude of the Cauchy stress within the third medium.}
    \label{fig:beaminbox}
\end{figure}

This example, intended to showcase the application of the third medium method to instability problems, is loosely inspired by the work of Puso and Solberg~\cite{puso2020contactbeaminbox}. Significant alterations had to be made to adapt the example to the third medium case, though. Figure~\ref{fig:beaminbox}a shows the geometry of this example, which consists of a rigid piston moving down through a rigid enclosed box, compressing a vertically oriented beam clamped to the piston on one end and to the bottom of the box on the other. To accommodate the sliding piston, the border nodes of the piston and third medium elements are not fixed to the rigid box and are only constrained horizontally instead (i.e., they can freely move vertically along the contact surface). The simulation has been performed up to the point at which the solver ceased to converge. A force-displacement diagram is shown in Figure~\ref{fig:beaminbox}b and successive deformed states of the example in Figures~\ref{fig:beaminbox}c-g. In the force-displacement diagram, the displacement is expressed in multiples of the piston and wall thickness $d$ and the force $f$ is presented in multiples of the Eulerian critical load $f_\mathrm{cr}$ for the double-clamped case:

\begin{equation}
    f_\mathrm{cr} = \frac{\pi^2 EI}{(0.5L)^2}
\end{equation}

\noindent where $L$ is the beam length between the clamps and $EI$ is its flexural rigidity. 

The corners in the force displacement diagram correspond to the various changes in configuration during the loading. First, the double-clamped beam buckles in response to the axial load (see the deformed configuration in Figure~\ref{fig:beaminbox}c). The buckling force corresponds to the Eulerian critical load within an error margin of $0.6\%$. The force then almost plateaus as the beam is compressed further in its buckled state, until contact is established with the side wall of the chamber, where the response stiffens again (Figure~\ref{fig:beaminbox}d). The stiffening is interrupted by the buckling of the axially compressed lower part of the beam (Figure~\ref{fig:beaminbox}e). This reduces the force until self contact is established late in the simulation (Figure~\ref{fig:beaminbox}f). Finally, the simulation ends after additional contact with the bottom of the box is established (Figure~\ref{fig:beaminbox}g, the ultimate converged solution step).

The parameters of the third medium were chosen to be somewhat stiffer here, with $\gamma_c = 10^{-6}$, $\gamma_e = 10^{-5}$, and $\kappa_{\bar{F}} = \SI{4000}{\pascal}$. We find this choice not to interfere with the results, as the observed contact gaps are minimal, perhaps with the exception of the states close to the simulation's end. Furthermore, the transfer of forces through the third medium is also assessed in the force-displacement diagram by showing the reaction forces measured both on the piston along its entire length, i.e., including also the interface with the third medium, and at the clamp of the beam at the bottom. Both plots overlap well until they detach upon establishment of contact between the beam and the bottom of the rigid box, which partially takes over the transfer of the vertical force. The deformed state figures show the magnitude of the Cauchy stress in the third medium, demonstrating how the contact forces are transferred through the compressed third medium elements. The upper bound of the color map has been capped at the 92nd percentile for better image clarity.

\subsection{Configurational forces}

This example is inspired by Dal Corso et~al.~\cite{dalcorso2024configforces}, who investigated an elastic solid undergoing large deformations when in frictionless contact with a rigid indenter. They showed that the energy release rate associated with an infinitesimal horizontal growth of the punch, the $J$-integral evaluated around its corner, and the horizontal component of the contact reaction force all coincide. Furthermore, they derived a closed-form analytical expression for the resulting configurational force, which was validated numerically using finite element simulations based on standard node-to-surface contact algorithms.

The goal of this example is to assess whether the configurational force so described can also be correctly recovered using the third medium approach. A key advantage of this benchmark is that Dal Corso et~al.~\cite{dalcorso2024configforces} provide a closed-form analytical solution, which offers a clear reference for validation. The analytical expression for the horizontal reaction force $R_1$ reads
\begin{equation}
    R_1 = \frac{W^r - W^l}{\lambda_1^r} \, h_0 ,
    \label{eq:configforce}
\end{equation}
where $W^l$ and $W^r$ are the strain energy density functions evaluated at the left and right boundaries of the solid, respectively, $h_0$ is the undeformed height of the solid, and $\lambda_1^r$ is the horizontal stretch measured at the right boundary.

\begin{figure}[t]
    \centering
    \includegraphics[width=.9\textwidth]{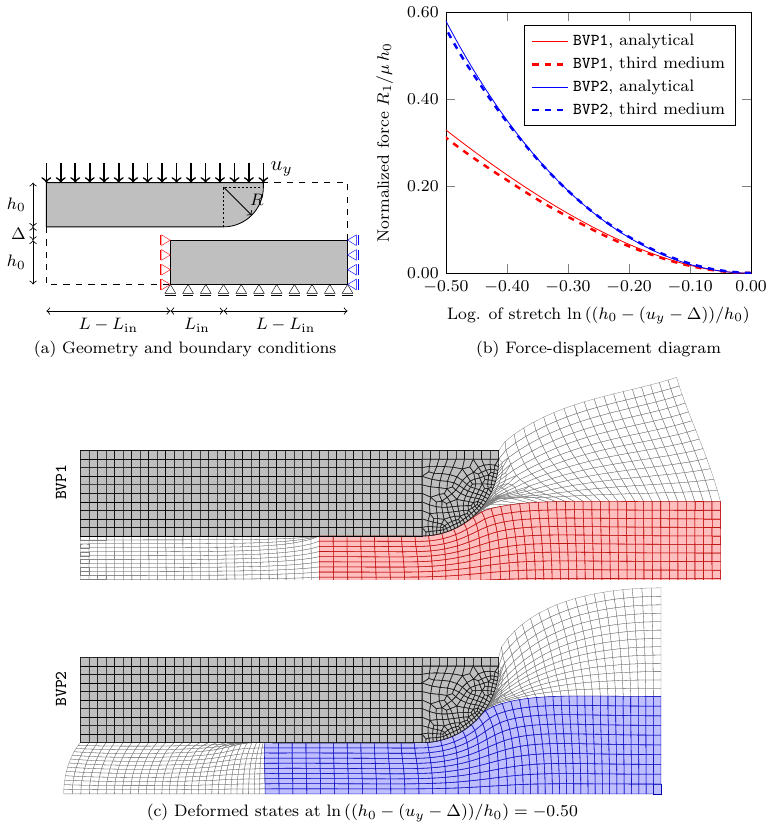}
    \caption{Configuration force problem. (a) Task geometry. (b) Force-displacement diagram for the two cases, $\mathtt{BVP1}$ and $\mathtt{BVP2}$, which differ by boundary conditions applied to the elastic block, denoted in (a) in red and blue, respectively. (c) Deformed states at the end of loading for the two cases.}
    \label{fig:bvp}
\end{figure}

Figure~\ref{fig:bvp}a describes the geometry of the two investigated boundary-value problems differing only in the location of the constraints applied to the elastic body. The boundary condition depicted in red corresponds to the boundary-value problem $\mathtt{BVP1}$, while that shown in blue corresponds to $\mathtt{BVP2}$. Dal Corso et~al.~\cite{dalcorso2024configforces} further showed that the radius of curvature of the indenter ($R$ in Figure~\ref{fig:bvp}a) does not affect the resulting configurational force, which depends only on the deformed state of the leftmost and the rightmost material fibers of the elastic block. This property makes the problems particularly suitable here, as it allows us to introduce a non-negligible value of $R$ and show how the third medium method handles the curved geometry of the contacting bodies while still being able to use the analytical solution for the configurational force as a benchmark. We chose $R = 0.9h_0$. The dimensions $L$, $L_\mathrm{in}$, and $\Delta$ (the initial contact gap) also do not influence the normalized analytical solution of the problem (see Equation~(\ref{eq:configforce})). In the presented simulations they were chosen as $L = 4h_0$, $L_\mathrm{in} = 0.3L$ and $\Delta = 0.3h_0$.

Figure~\ref{fig:bvp}b shows the evolution of the dimensionless horizontal reaction force $R_1$ with respect to the logarithmic vertical stretch of the right boundary of the solid. Results are shown for $\mathtt{BVP1}$ (red) and $\mathtt{BVP2}$ (blue), where the analytical solutions (solid lines) are compared with the third medium results (dashed lines). An excellent agreement between the analytical predictions and the third medium formulation is observed for both boundary-value problems over the entire loading range.

The final deformed configurations for both boundary-value problems are presented in Figure~\ref{fig:bvp}c. The results illustrate how the third medium adapts to the deformation and effectively compresses close to zero volume along the contact interfaces once contact is established. Due to the large deformations involved, further loading leads to sliding of the elastic solid along the indenter, subjecting the third medium not only to vertical compression but also to significant shear deformation. The ability to accurately reproduce the configurational force in this regime demonstrates the robustness of the proposed third medium formulation under combined normal and tangential interactions.

The results presented in this section are obtained using the parameters $\gamma_\mathrm{a} = 10^{-6}$, $\gamma_\mathrm{b} = 10^{-3}$, and $\kappa_{\bar{F}} = 2\cdot10^{5}\;\si{\pascal}$. The elastic solid is described by an incompressible neo-Hookean model in agreement with the examples presented by Dal~Corso~et~al.~\cite{dalcorso2024configforces}, with $E=3\cdot10^6\;\si{\pascal}$ and $\nu = 0.49$. In contrast to the previous examples, here the mesh consists of 8-node quadrilateral and 6-node triangular finite elements with quadratic shape functions. The standard Gauss integration rule is employed, with 16 integration points per element. This is motivated by the desire to give the third medium elements located in the gap more kinematic freedom, since they are required to slide and bend around the curved contact surface.

\section{Conclusions}
\label{sec:conclusions}

In this contribution, we introduced a third medium contact model which, in contrast to all existing models, requires only a first-order finite element formulation without additional degrees of freedom. There are two key components: a small-strain-inspired constant-stiffness law for the third medium and a regularization based on deformation averaging.

Moreover, we use a small-strain-like constant-stiffness law for the third medium in the pre-contact phase that keeps the stiffness of the third medium effectively constant throughout the simulation with the exception of contact enforcement. Its inclusion allowed us to limit the neo-Hookean contribution, which typically forms the basis of third medium models, to its volumetric term only, which enforces contact by reacting asymptotically to extreme compression. The isochoric term of the neo-Hookean elasticity was omitted, which improves the stability of the  highly compliant third medium elements.

The deformation-averaging term, which penalizes non-uniformity of the deformation gradient tensor across the element integration points, was motivated by the similarity of this approach to the element-wise penalization of second displacement gradients commonly found in third medium contact models. This usual penalization, however, requires at least second-order element formulation, or alternative treatments based on extra degrees of freedom. In our approach, standard first-order element formulation is sufficient, even though the regularization is only meaningful when multiple integration points with theoretically different values of the deformation gradient tensor exist within the element. Thus, it leads to a very simple formulation.

We expect this new approach to find use in third medium contact simulations of both two-dimensional and three-dimensional settings. It is easy to implement, universal, and robust, as we have demonstrated through multiple examples in this work. All components of the third medium model admit an energy potential, which is a useful property for advanced finite element solvers, including solvers for stability problems. As with previous third medium models, we expect our model to be useful to the topology optimization community. Likewise, the regularization techniques developed here can find use in a range of multiphysical finite element simulations in which air-like space needs to be meshed.

\paragraph{Acknowledgement}
This work has been supported by the  Ministry of Education, Youth and Sport of the Czech Republic (ERC CZ project no. LL2310).

\appendix
\setcounter{figure}{0}   
\section{Implementation of the constant stiffness term}
\label{app:isole_details}

In Section~\ref{sec:constantstiffnessterm} we propose the use of a constant stiffness linear elastic material law for the third medium. Its strain energy density, expressed in terms of the deformation gradient $\bm{F}$, is given as

\begin{equation}
\label{appeq:smallstrainenergydensity}
    W_\mathrm{ss} = \frac{1}{2}\bm{\varepsilon} : \mathbf{D}^\mathrm{3M} : \bm{\varepsilon} = \frac{1}{2} \left(\bm{F}-\bm{I}\right) : \mathbf{D}^\mathrm{3M} : \left(\bm{F}-\bm{I}\right)
\end{equation}

\noindent where $\bm{\varepsilon} = \nabla_\mathrm{s}\bm{u}$ is the engineering strain, i.e., a symmetric gradient of the displacement field, $\mathbf{D}^\mathrm{3M}$ is the small strain material stiffness tensor for the third medium, and $\bm{F} = \bm{I} + \nabla\bm{u}$ is the deformation gradient.

The second equality in Equation~(\ref{appeq:smallstrainenergydensity}) uses the minor symmetries of the material stiffness tensor, defined in index notation as

\begin{equation}
    D^\mathrm{3M}_{ijkl} = \lambda \delta_{ij}\delta_{kl} + \mu \left(\delta_{ik}\delta_{jl} + \delta_{il}\delta_{jk}\right)
\end{equation}

\noindent with $\lambda$ and $\mu$ as the Lamé coefficients. In the term $\mathbf{D}^\mathrm{3M} : \left(\bm{F}-\bm{I}\right)$ the displacement field gradient can be divided into its symmetric and anti-symmetric parts as

\begin{equation}
\label{appeq:symandantisymDtimesF}
    D^\mathrm{3M}_{ijkl} \left(F_{kl} - \delta_{kl}\right) = \frac{1}{2} D^\mathrm{3M}_{ijkl} \left(u_{k,l} + u_{l,k}\right) + \frac{1}{2} D^\mathrm{3M}_{ijkl} \left(u_{k,l} - u_{l,k}\right) 
\end{equation}

\noindent with the notation $u_{a,b}$ referring to the gradient operation $\partial u_a/\partial X_b$. It then follows from the right minor symmetry of $\mathbf{D}^\mathrm{3M}$ that

\begin{equation}
     D^\mathrm{3M}_{ijkl} u_{l,k} = D^\mathrm{3M}_{ijlk} u_{l,k} = D^\mathrm{3M}_{ijkl} u_{k,l}
\end{equation}

\noindent and therefore the last term in Equation~(\ref{appeq:symandantisymDtimesF}) vanishes, leaving

\begin{equation}
    D^\mathrm{3M}_{ijkl} \left(F_{kl} - \delta_{kl}\right) = \frac{1}{2} D^\mathrm{3M}_{ijkl} \left(u_{k,l} + u_{l,k}\right) = D^\mathrm{3M}_{ijkl} \varepsilon_{kl}
\end{equation}

\noindent A similar argument can be made using the left minor symmetry of $\mathbf{D}^\mathrm{3M}$ for the term $\left(\bm{F}-\bm{I}\right) : \mathbf{D}^\mathrm{3M}$.

We can construct the first and second variation of the strain energy density (\ref{appeq:smallstrainenergydensity}) as

\begin{eqnarray}
    \delta W_\mathrm{ss} &=&  \delta \bm{F} : \mathbf{D}^\mathrm{3M} : \left(\bm{F}-\bm{I}\right)
    \\
    \delta^2 W_\mathrm{ss} &=& \delta \bm{F} : \mathbf{D}^\mathrm{3M} : \delta \bm{F}
\end{eqnarray}

The definition of the first Piola-Kirchhoff stress $\bm{P}$ follows as

\begin{equation}
    \bm{P} = \mathbf{D}^\mathrm{3M} : \left(\bm{F}-\bm{I}\right)
\end{equation}

\noindent while the material stiffness tensor is indeed just the constant tensor $\mathbf{D}^\mathrm{3M}$. Due to similar arguments as above, the first Piola-Kirchhoff stress is always symmetric here.

\section{Implementation of the deformation averaging regularization}
\label{app:fbar_details}

For deformation averaging regularization the strain energy density of the third medium is enhanced by the term
\begin{equation}
    W_{\bar{F}} = \dfrac{1}{2}\kappa_{\bar{F}} \|\bm{F}-\bar{\bm{F}}\|^2
\label{appeq:barFenergy}
\end{equation}

\noindent where $\kappa_{\bar{F}}$ is a penalty parameter, $\bm{F}$ is the deformation gradient, $\bar{\bm{F}}$ is the deformation gradient at the finite element centroid, and $\|\bm{a}\|$ signifies the Frobenius norm of a tensor $\bm{a}$.

Thus, the first variation of the strain energy density function is 

\begin{equation}
    \delta W_{\bar{F}} = \underbrace{\kappa_{\bar{F}}\left(\bm{F}-\bar{\bm{F}}\right)}_{\bm{P}_{\bar{F}}}:\delta \bm{F} + \underbrace{\kappa_{\bar{F}}\left(-\bm{F}+\bar{\bm{F}}\right)}_{\bar{\bm{P}}_{\bar{F}}}:\delta \bar{\bm{F}}
\label{appeq:barFenergyFirstVar}
\end{equation}

\noindent defining the first Piola-Kirchhoff stress tensor $\bm{P}_{\bar{F}} = \partial W_{\bar{F}}/\partial \bm{F}$ and its center-of-element counterpart $\bar{\bm{P}}_{\bar{F}} = \partial W_{\bar{F}}/\partial \bar{\bm{F}}$. Similarly the second variation of strain energy density is

\begin{eqnarray}
\label{appeq:barFenergySecondVar}
    \delta^2 W_{\bar{F}} &=& \delta \bm{F} : \underbrace{\left(\kappa_{\bar{F}}\mathbf{I}\right)}_{\mathbf{D}^1_{\bar{F}}}:\delta \bm{F} + \delta \bm{F} : \underbrace{\left(-\kappa_{\bar{F}}\mathbf{I}\right)}_{\mathbf{D}^2_{\bar{F}}}:\delta \bar{\bm{F}}
    \\ \nonumber
    &+& \delta\bar{\bm{F}} : \underbrace{\left(-\kappa_{\bar{F}}\mathbf{I}\right)}_{\mathbf{D}^3_{\bar{F}}}:\delta \bm{F} + \delta \bar{\bm{F}} : \underbrace{\left(\kappa_{\bar{F}}\mathbf{I}\right)}_{\mathbf{D}^4_{\bar{F}}}: \delta\bar{\bm{F}}
\end{eqnarray}

\noindent with $\mathbf{I}$ being the fourth order identity tensor, where the four stiffness terms $\mathbf{D}^1_{\bar{F}}$ to $\mathbf{D}^4_{\bar{F}}$ are defined as the partial derivatives of $\bm{P}_{\bar{F}}$ and $\bar{\bm{P}}_{\bar{F}}$ with respect to either $\bm{F}$ or $\bar{\bm{F}}$.

Transferred to index notation, Equation~(\ref{appeq:barFenergy}) reads

\begin{equation}
    W_{\bar{F}} = \dfrac{1}{2}\kappa_{\bar{F}}\left(F_{ij}-\bar{F}_{ij}\right)\left(F_{ij}-\bar{F}_{ij}\right)
\end{equation}

\noindent The stresses are then defined as 
\begin{eqnarray}
    P_{ij} &=& \dfrac{\partial W}{\partial F_{ij}} = \kappa_{\bar{F}}\left(F_{ij}-\bar{F}_{ij}\right)
    \\
    \bar{P}_{ij} &=& \dfrac{\partial W}{\partial \bar{F}_{ij}} = -\kappa_{\bar{F}}\left(F_{ij}-\bar{F}_{ij}\right)
\end{eqnarray}
\noindent since only the contributions $\bm{P}_{\bar{F}}$ and $\bar{\bm{P}}_{\bar{F}}$ are considered in this section, the subscript $\bar{F}$ is omitted in the index notation for brevity. The stiffness tensors are defined  as
\begin{eqnarray}
    D^1_{ijkl} &=& \dfrac{\partial P_{ij}}{\partial F_{kl}} = \kappa_{\bar{F}} \, \delta_{ik} \, \delta_{jl}
    \\
    D^2_{ijkl} &=& \dfrac{\partial P_{ij}}{\partial \bar{F}_{kl}} = -\kappa_{\bar{F}} \, \delta_{ik} \, \delta_{jl}
\end{eqnarray}
\begin{eqnarray}
    D^3_{ijkl} &=& \dfrac{\partial \bar{P}_{ij}}{\partial F_{kl}} = -\kappa_{\bar{F}} \, \delta_{ik} \, \delta_{jl}
    \\
    D^4_{ijkl} &=& \dfrac{\partial \bar{P}_{ij}}{\partial \bar{F}_{kl}} = \kappa_{\bar{F}} \,\delta_{ik} \, \delta_{jl}
\end{eqnarray}

Upon finite element discretization, the displacement field $\bm{u}$ within a finite element is approximated by the product of the element shape function matrix $\mathsf{N}$ and the vector of nodal displacements $\mathsf{d}$ as $\bm{u}(\bm{X}) \approx \mathsf{N}\big\rvert_{\bm{\xi} = \bm{\xi}_X} \mathsf{d}$, with $\bm{X}$ being the material coordinate vector, $\bm{\xi}$ the element natural coordinate vector, and $\bm{\xi}_X$ its value at a point corresponding to $\bm{X}$. Similarly the variation of the displacement field $\delta\bm{u}$ is approximated as $\delta\bm{u}(\bm{X}) \approx \mathsf{N}\big\rvert_{\bm{\xi} = \bm{\xi}_X} \delta\mathsf{d}$.

Equations~(\ref{appeq:barFenergyFirstVar}) and (\ref{appeq:barFenergySecondVar}) also transform, in Voigt notation, to

\begin{eqnarray}
\label{appeq:barFenergyFirstVarVoigt}
    \delta W_{\bar{F}} &=& \mathsf{P}^T\delta \mathsf{F} + \bar{\mathsf{P}}^T\delta \bar{\mathsf{F}}
    \\ \label{appeq:barFenergySecondVarVoigt}
    \delta^2 W_{\bar{F}} &=& \delta \mathsf{F}^T \mathsf{D}^1 \delta \mathsf{F} + \delta \mathsf{F}^T \mathsf{D}^2 \delta \bar{\mathsf{F}} + \delta\bar{\mathsf{F}}^T \mathsf{D}^3 \delta \mathsf{F} + \delta \bar{\mathsf{F}}^T \mathsf{D}^4 \delta\bar{\mathsf{F}}
\end{eqnarray}

\noindent where $\mathsf{P}$, $\mathsf{\bar{P}}$, $\mathsf{F}$, $\mathsf{\bar{F}}$, $\mathsf{D}^1$, $\mathsf{D}^2$, $\mathsf{D}^3$, and $\mathsf{D}^4$ were introduced as the Voigt representations of the tensors $\bm{P}$, $\bm{\bar{P}}$, $\bm{F}$, $\bm{\bar{F}}$, $\mathbf{D}^1$, $\mathbf{D}^2$, $\mathbf{D}^3$, and $\mathbf{D}^4$, respectively.

The domain-wise integration of Equations~(\ref{appeq:barFenergyFirstVar}) and (\ref{appeq:barFenergySecondVar}) is then superseded by the sum of element-wise integrations of (\ref{appeq:barFenergyFirstVarVoigt}) and (\ref{appeq:barFenergySecondVarVoigt}). Those element-wise integrations are themselves performed as a sum over integration points:

\begin{eqnarray}
    \int_{\Omega_\mathrm{e}} \delta W_{\bar{F}} \; \mathrm{d}\Omega &\approx& \sum_{i_\mathrm{g}=1}^{n_\mathrm{g}} \alpha_i \left( \mathsf{P}^T\delta \mathsf{F} + \bar{\mathsf{P}}^T\delta \bar{\mathsf{F}} \right)
    \\
    \int_{\Omega_\mathrm{e}} \delta^2 W_{\bar{F}} \; \mathrm{d}\Omega &\approx& \sum_{i_\mathrm{g}=1}^{n_\mathrm{g}} \alpha_i \left(\delta \mathsf{F}^T \mathsf{D}^1 \delta \mathsf{F} + \delta \mathsf{F}^T \mathsf{D}^2 \delta \bar{\mathsf{F}} + \delta\bar{\mathsf{F}}^T \mathsf{D}^3 \delta \mathsf{F} + \delta \bar{\mathsf{F}}^T \mathsf{D}^4 \delta\bar{\mathsf{F}}\right)
\end{eqnarray}

\noindent where $\Omega_\mathrm{e}$ denotes the area of a finite element, $n_\mathrm{g}$ is the number of its Gauss integration points, and $\alpha_i$ is the integration weight of the $i$-th integration point. In this context the deformation gradient $\mathsf{F}$ and its variation $\delta\mathsf{F}$ are approximated at a given Gauss point using the matrix of shape function gradients $\mathsf{B}$ evaluated at the Gauss point's natural coordinates $\mathsf{\xi}_\mathrm{g}$ as

\begin{equation}
    \mathsf{F} \approx \mathsf{I} + \mathsf{B}\big\rvert_{\xi = \xi_\mathrm{g}} \mathsf{d} \quad\quad\quad\quad \delta\mathsf{F} \approx \mathsf{B}\big\rvert_{\xi = \xi_\mathrm{g}}\delta\mathsf{d}
\end{equation}

\noindent where $\mathsf{I}$ is the identity matrix. The finite element approximation of $\bar{\mathsf{F}}$, on the other hand, can be obtained by simply evaluating the shape function gradients always at the element centroid, e.g., at $\xi_0 = (0,0)$, which corresponds to the centroid in the natural coordinate system of standard isoparametric two-dimensional elements:

\begin{equation}
    \bar{\mathsf{F}} \approx \mathsf{I} + \mathsf{B}\big\rvert_{\xi = \xi_0} \mathsf{d} \quad\quad\quad\quad \delta\bar{\mathsf{F}} \approx \mathsf{B}\big\rvert_{\xi = \xi_0}\delta\mathsf{d}
\end{equation}

\noindent Moreover, the element force vector $\mathsf{f}_\mathrm{e}$ and element stiffness matrix $\mathsf{K}_\mathrm{e}$ are obtained by cancelling the variational nodal displacement vector $\delta \mathsf{d}$ from Equation~(\ref{appeq:barFenergyFirstVarVoigt}) and Equation~(\ref{appeq:barFenergySecondVarVoigt}), respectively, leaving

\begin{eqnarray}
    \mathsf{f}_\mathrm{e} &=& \sum_{i_\mathrm{g}=1}^{n_\mathrm{g}}\alpha_i\left( \mathsf{B}^T \mathsf{P} + \bar{\mathsf{B}}^T\bar{\mathsf{P}}\right)
    \\
    \mathsf{K}_\mathrm{e} &=& \sum_{i_\mathrm{g}=1}^{n_\mathrm{g}}\alpha_i\left( \mathsf{B}^T \mathsf{D}^1 \mathsf{B} + \mathsf{B}^T \mathsf{D}^2 \bar{\mathsf{B}} + \bar{\mathsf{B}}^T \mathsf{D}^3 \mathsf{B} + \bar{\mathsf{B}}^T \mathsf{D}^4 \bar{\mathsf{B}}\right)
\end{eqnarray}

\noindent where, for brevity, the symbol $\bar{\mathsf{B}}$ has been introduced to denote the $\mathsf{B}$ matrix evaluated at $\xi_0$.

\bibliographystyle{elsarticle-num} 
\bibliography{biblio2.bib}





\end{document}